\documentclass{article}

\usepackage{arxiv}
\usepackage{xcolor}
\usepackage{chngcntr}
\counterwithin{figure}{section}
\counterwithin{equation}{section}

\newcommand{\beginsupplement}{
        \setcounter{table}{0}
        \renewcommand{\thetable}{S\arabic{table}}
        \setcounter{figure}{0}
        \renewcommand{\thefigure}{S\arabic{figure}}
        \setcounter{figure}{0}
        \renewcommand{\thefigure}{S\arabic{figure}}
        \setcounter{section}{1}
        \renewcommand{\thesubsection}{S\arabic{subsection}}
        \setcounter{equation}{0}
        \renewcommand{\theequation}{S\arabic{equation}}
}

\usepackage[utf8]{inputenc} 
\usepackage[T1]{fontenc}    
\usepackage{hyperref}       
\usepackage{url}            
\usepackage{booktabs}       
\usepackage{amsfonts}       
\usepackage{nicefrac}       
\usepackage{microtype}      
\usepackage{lipsum}
\usepackage{amsmath}
\usepackage{graphicx}
\usepackage{placeins}
\usepackage{float}
\usepackage{comment}

\title{Reconstruction of Pairwise Interactions using Energy-Based Models}

\author{
  Christoph Feinauer\\
  Department of Decision Sciences \\
  Bocconi Institute for Data Science and Analytics (BIDSA)\\
  Bocconi University, Milan, Italy\\
  \texttt{christoph.feinauer@unibocconi.it} \\
  \And{}
  Carlo Lucibello \\
  Department of Decision Sciences \\
  Bocconi Institute for Data Science and Analytics (BIDSA)\\
  Bocconi University, Milan, Italy\\
  \texttt{carlo.lucibello@unibocconi.it} \\
}

\begin{document}
\maketitle

\begin{abstract}
    Pairwise models like the Ising model or the generalized Potts model have
    found many successful applications in fields like physics, biology and
    economics. Closely connected is the problem of inverse statistical
    mechanics, where the goal is to infer the parameters of such models given
    observed data. An open problem in this field is the question on how to
    train these models in the case where the data contain additional
    higher-order interactions that are not present in the pairwise model. In
    this work, we propose an approach based on Energy-Based Models
    and pseudolikelihood maximization to
    address these complications: we show that hybrid models, which combine a
    pairwise model and a neural network, can lead to significant improvements in
    the reconstruction of pairwise interactions. We show these improvements to
    hold consistently when compared to a standard approach using only the
    pairwise model and to an approach using only a neural network.  This is in
    line with the general idea that simple interpretable models and complex
    black-box models are not necessarily a dichotomy: interpolating these two
    classes of models can allow to keep some advantages of both.
\end{abstract}

\keywords{Pairwise Models, Neural Networks, Inverse Ising, Energy Based Models}

\section{Introduction}

An important class of distributions used in the modeling of natural systems is
the exponential family of pairwise models. Commonly investigated in the
statistical physics community, pairwise models are a popular method for the
analysis of categorical sequence data. Examples of data on which they have been
successfully applied include protein sequence data~\cite{Morcos2011,
marks2012protein, Cocco2018}, neuronal recordings~\cite{roudi2009ising,
tkavcik2014searching}, magnetic spins~\cite{fisher1986ordered}, economics and
social networks~\cite{stauffer2008social, sornette2014physics,
hall2019statistical}.

One main advantage of these models is their relative simplicity: The
probability assigned to a sequence $s$ of binary or categorical variables is of
the form $p(s) \propto \exp(-E(s))$, where $E$ is a \textit{simple} function of
$s$, meaning that it consists of terms that depend on only one or two
variables. The parameters quantifying the pairwise interactions are
typically called \textit{couplings}.

Given this simple form, the parameters can often be given a direct
interpretation in terms of the underlying system. Especially the couplings have
been shown to contain highly non-trivial information in many cases: The
couplings in the so-called Potts Models for protein sequence data can be seen
as a measure for the strength of co-evolutionary pressure between parts of the
sequence and can be used for the prediction of structural
features~\cite{Morcos2011}; the couplings in models for neuronal recordings can
be seen as the functional couplings between neurons~\cite{roudi2009ising}; the
couplings in magnetic systems of interacting spins can be seen as describing
their physical interaction strength.

While pairwise models have been surprisingly successful in many fields, they
have clear limitations: If the data generating process contains important
interactions that cannot be described as pairwise interactions, the models
might fail to capture important variability. Even worse, if such interactions
are strong enough, the pairwise models might even stop to describe the pairwise
interactions properly since they might contain effective pairwise interactions that
try to include the variability of the higher-order interactions. In fact, it is
known in literature that for example some variability in protein sequences is due to
higher-order epistasis, including more than 2 residues~\cite{Waechter2012}.

Several methods have been proposed to address such problems, for example the
`manual' addition of higher-order interaction terms based on a close look at
the data~\cite{Feinauer2014}, or the addition of complete sets of higher-order
interactions, for example all terms involving triplets of
variables~\cite{schmidt2018hodca}.

Another option is to abandon simple pairwise models and adopt more complicated,
but also more expressive methods, for example neural networks. These models can
in principle capture interactions of all orders and can be trained for a
specific task, for example the extraction of structural information
~\cite{peng2011raptorx}, the generation of new samples
~\cite{riesselman2019accelerating} or the creation of generic embeddings
~\cite{rives2019biological}. While this strategy has lead to unprecedented successes in many fields, it also
comes at the cost of a higher computational demand and the loss of the
interpretability of the single parameters defining the distribution. Moreover,
failing to encode the prior knowledge on the data generative process, 
these black-box methods are far away from being optimal in terms of sample efficiency.

In this paper, we propose a combination of these two approaches: We develop
a strategy for keeping the simple model with its advantages in place, but add a
neural network model to \textit{help} with the more complex patterns in the
data. This approach seems sensible in cases where we suspect or know that a
simple model is able to capture most of the variability in the data, but that
it might fail to capture some additional aspects or even gets confused by them.

We implement this idea defining a new energy function

\begin{align}
    E(s) = E_{pw}(s) + E_{nn}(s),
\end{align}

where $E_{pw}$ is a pairwise model and $E_{nn}$ is a neural network that maps a
configuration $s$ to a real number. We then look at cases where the data generating
process contains a simple part, corresponding to another pairwise model, and a
more complicated part, corresponding to higher-order interactions. The hope is
that the neural network picks up these higher-order interactions and thus helps
the pairwise model in matching the pairwise interactions of the generative process.

We will focus on the so-called inverse problem of statistical mechanics, that is reconstructing the pairwise couplings of a generative model containing also some unknown higher-order interaction terms. 

\section{Methods}
\label{sec:methods}
\subsection{Pairwise Models and Energy-Based Models}
We consider a probability distribution $p(s)$ over all possible configurations of $N$ binary variables, $\{-1,+1\}^N$. Any such distribution with support over the whole space can be written in the form

\begin{align}
    p(s) = \exp(-E(s))/Z,
\end{align}

where $E: s \rightarrow \mathcal{R}$ is the so called energy function and $Z$ is a normalization constant called the partition function.
Denoting, with $\mathcal{I}$ the power set of $\{1,\ldots,N\}$, the energy can be uniquely expressed by the expansion

\begin{align}
    E(s) = - \sum_{I \in \mathcal{I}} \xi_{I} \prod_{i \in I} s_i,
    \label{eq:energy_expansion}
\end{align}

where $\xi_{I} \in \mathcal{R}$ is the interaction coefficient for the term
containing the variables specified by $I$.  Such expansions are known in
theoretical computer science and Boolean algebra as \textit{Fourier
expansions}, and the corresponding parameters $\xi_{I}$ are called
\textit{Fourier coefficients} \cite{o2014analysis}. Determining specific
coefficients from a black-box function $E$ can be done efficiently through
sampling techniques (see Section \ref{sec:extraction}) and coefficients larger
than a given threshold can be determined using the \textit{Goldreich-Levin}
algorithm~\cite{o2014analysis}. This is useful in our setting, since these
techniques also apply when the energy $E$ is parametrized using an arbitrary
neural network.

%
%
%

The class of models where $\xi_{I}=0$ if $|I|>2$ are called pairwise models, defined by the energy
\begin{equation}
E_{pw}(s) = - \sum\limits_{i} h_i s_i - \sum_{i<j} J_{ij} s_i s_j.    
\end{equation}
The coefficients $h_i$ are called external fields and the coefficients $J_{ij}$ are called couplings. Such models have a long history of statistical physics and have been exported
to various fields. In a typical application, the model is fitted to a dataset
$D=\{s^{m}\}_{m=1}^{M}$ consisting of $M$ configurations sampled from the system
under study, and can be afterwards either used as a generative model or 
insights about the system can be gained from examining the fitted parameters $J$ and
$h$.  However, if we assume the existence of a generating distribution
$p_{G}(s)$ that includes important interactions involving more than two
variables, the pairwise distribution might fail to describe the variability in
the dataset (see Section~\ref{sec:results}) and the inferred couplings and fields might not
correspond to the ones in the generating distribution. If such a case is suspected, one is tempted to use a more complicated function
to describe the energy $E$. Given the flexibility of neural
networks in approximating arbitrary input-output dependencies, a promising choice could be a multi-layer perceptron with $L$ layers, where the operations in each layer are a matrix multiplication followed by the addition of a bias and the application of a possibly non-linear  activation function~\cite{goodfellow2016deep}.

Models that use a neural network for parameterizing the energy are commonly
called \textit{Energy-Based Models} (EBMs)~\cite{lecun2006tutorial} and have
been applied recently with success for example in the field of image generation
~\cite{du2019implicit}. Apart from the appealing similarity to models used in statistical mechanics
since more than a hundred years, they present several advantages in comparison
to other model classes like \textit{Generative Adversarial
Networks}~\cite{goodfellow2014generative} or \textit{Variational
Autoencoders}~\cite{kingma2013auto}. The most important ones related to the
present work are their relative uniformity and simplicity and their
composionality (both also mentioned in~\cite{du2019implicit}). By uniformity
and simplicity we refer to the fact that due to the generic formulation using a
single energy function, tasks like training, sampling and analysis can often
be formulated generically, independent of the exact parametrization of the
energy.  By composionality, we refer to the idea that EBMs can be combined
easily by summing their respective energies, leading to a so called
\textit{product of experts} model~\cite{hinton2002training}. In fact, the
central idea of this paper is to combine two different energy functions: One
simple, and one complex. The main disadvantage of EBMs is that the
normalization constants appearing in the definition of the probability cannot
be calculated for typical input sizes, which makes training and sampling
harder. However, many training methods like \textit{contrastive
divergence}~\cite{carreira2005contrastive} or \textit{pseudolikelihoods} (see
SM Section~\ref{sec:pl_sup}) can be adapted for EBMs,
and sampling can be done using standard MCMC algorithms like
\textit{Metropolis-Hastings}~\cite{metropolis1953equation}.

While such models are in principle capable of fitting
any probability distribution with increasing layer sizes, the amount of data
and the computational resources needed are typically much larger than for
fitting a pairwise model. In addition, they might not be very efficient: In the
trivial case that the generating distribution is in fact a pairwise
distribution, the number of parameters that need to be introduced to the EBM
might greatly exceed the ones needed when fitting a simpler distribution.

\subsection{Hybrid Models and Extraction of Coefficients}
\label{sec:extraction}

\begin{figure}
    \centering
    \includegraphics[width=0.99\textwidth]{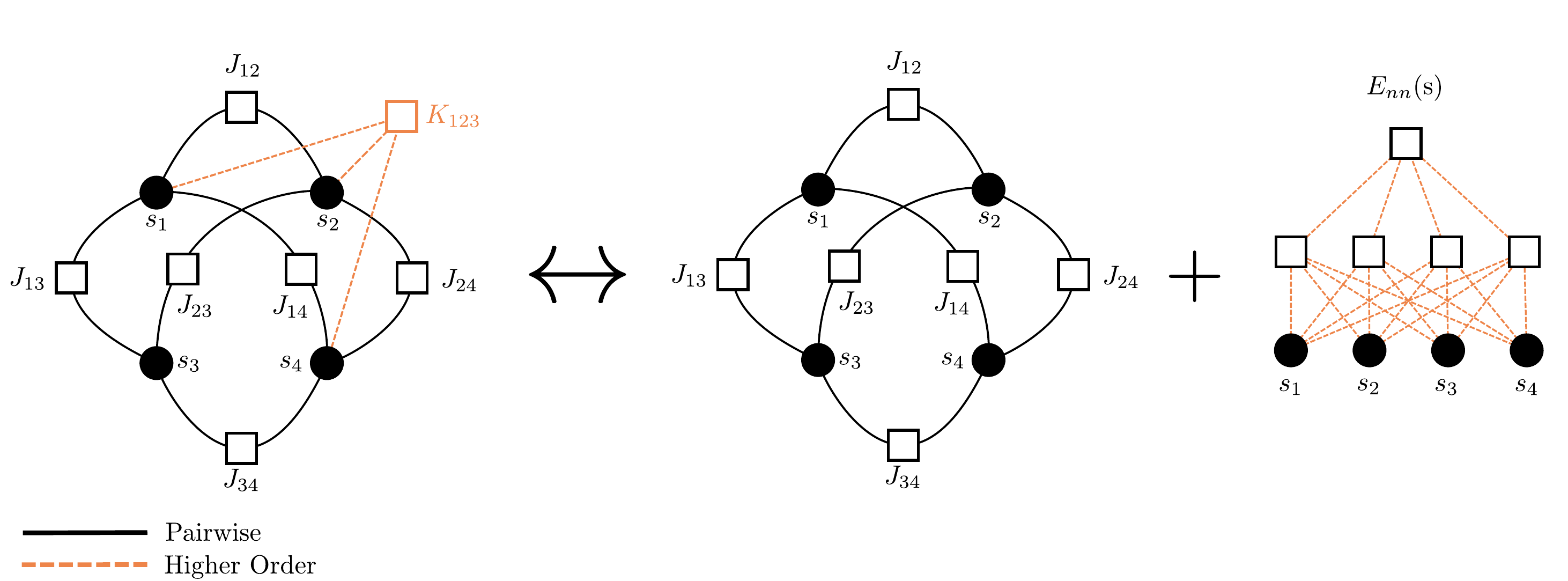}
    \caption{Representation of the basic idea of this work: Given a generative distribution that contains a strong pairwise part but also higher-order interactions, we fit an energy-based model including a pairwise part and a neural network. The hope is that the neural network captures the higher-order interactions, while the pairwise parts match up after training.}
    \label{fig:diagram}
\end{figure}

The models we use in this work are hybrid models of the form

\begin{align}
    E(s) = E_{pw}(s) + E_{nn}(s) =  -\sum\limits_{i<j} J_{ij} s_i s_j + E_{nn}(s).
    \label{eq:hybrid}
\end{align}

For simplicity and since we want to focus on the more complex problem of reconstructing the couplings, we do not explicitly consider external fields $h_i$ in this work, although they could be easily accounted for. $E_{nn}(s)$ is a
neural network with one hidden layer with $\tanh$ activations. While we could also test networks with more than one layer, there is evidence that in similar settings the most important characteristic is still the size of the first hidden layer, while the depth is of minor importance \cite{morningstar2017deep}. Since adding depth would also add the problem of finding the optimal architecture, we restrict ourselves to a single layer in this work, and also leave the exploration of other methods like self-attention \cite{vaswani2017attention} or autoregressive architectures \cite{wu2019solving} for further research.

One interpretation of
these models is that we model the pairwise terms in the expansion
Eq.~\eqref{eq:energy_expansion} explicitly, while we use a neural network for
describing all other interactions (see Fig. \ref{fig:diagram}).  For the neural network part $E_{nn}$, the
general expansion in Eq.  \eqref{eq:energy_expansion} can contain in principle
interactions of all orders.  For small system sizes, we can extract the
corresponding interaction parameters by observing from Eq.
\eqref{eq:energy_expansion} that for a generic energy $E(s)$ we have

\begin{equation}
\xi_{I} = -\mathbb{E}_{s} \left[ \ E(s) \prod_{i \in I} s_i \right],
\label{eq:interaction_estimate}
\end{equation}
where the expectation is according to the uniform distribution over all possible $2^N$ configurations.
Since we do not limit the capacity of the neural network, $E_{nn}(s)$ can also
contain significant pairwise interactions. Therefore, we may have $  \mathbb{E}_{s} [ E(s)s_i s_j ] \not\approx -J_{ij} $.  We show below that this can be indeed
observed in specific situations and approach the problem as follows:  we
reconstruct the couplings from $E=E_{pw} + E_{nn}$ using Eq. \eqref{eq:interaction_estimate}. We
refer to these effective couplings as \textit{reconstructed} couplings
\begin{align}
    \hat{J}_{ij}  = -\mathbb{E}_s \left[ E(s)\,s_i s_j \right],
\label{eq:coupling_estimate}
\end{align}
as opposed to the \textit{explicit} couplings $J$ in the trained model \eqref{eq:hybrid}.
The reconstruction is performed only at the end of the training, and approximated for large systems with $10^{6}$ Monte Carlo samples in our experiments. As an alternative, in SM Section~\ref{sec:absorber}, we show that it can be also done during training, which effectively limits the pairwise interactions in the hybrid model to the pairwise part. 

We use the same reconstruction method for extracting coefficients in models consisting \textit{only} of the neural network, without the explicit pairwise part, to
understand whether using an explicit pairwise term in model \eqref{eq:hybrid} brings any advantage. While
we do this here only for comparison and use only simple multi-layer
perceptrons (MLP), we note that it would be an interesting avenue of research to use
more advanced neural network models and see if the extracted couplings can be
used in applications where pairwise models are typically used.

\subsection{Training Procedure}
\label{sec:training}

The difficulties in evaluating the normalization constant in energy-based models make density evaluation intractable, and efficient
sampling becomes problematic as well. Many techniques have been proposed for the
challenging task of training EBMs, the most commonly used ones being contrastive
divergence with Langevin dynamics~\cite{hinton2002training,Du2019},
noise-contrastive estimation~\cite{gutmann2010noise}, and score matching
\cite{hyvarinen2005estimation}. In this work, we use pseudolikelihood
maximization to train the parameters of the model given the
data~\cite{besag1977efficiency}.  This method is very popular for the training
of pairwise models~\cite{aurell2012inverse, ekeberg2013improved, Decelle2014}
and is furthermore very similar to the method of training for state-of-the-art
neural network models summarily called \textit{self-supervised learning}, which
transforms the task of unsupervised learning of unlabeled data into a
supervised learning task by training the model to predict an artificially
masked part of the data from the  unmasked part. This technique is for example
used when training the self-attention based \textit{Bert}
models~\cite{devlin2018bert}. 
Given a single mini-batch $\{s^b\}_{b=1}^B$ with $B$ training configurations, we use the negative pseudo-likelihood loss function

\begin{align}
    \mathcal{L} = -\frac{1}{B} \sum\limits_{b=1}^B \sum\limits_{i=1}^{N} \log p(s^b_i | s^b_{/i}),
\end{align}

where the quantity $p(s^b_i | s^b_{/i})$ corresponds to the
probability of observing $s^b_i$ given the other variables in $s^b$, excluding
$s^b_i$. This loss function can be calculated for a generic energy model over configurations using $2N$ forward passes. For a
pairwise model instead, we can use more efficient calculation schemes. For implementation details see Supplementary \ref{sec:pl_sup}.
It is worth mentioning that the interaction screening approach of Ref. \cite{vuffray2020efficient} provides an alternative with well understood sample complexity guarantees to the pseudolikelihood framework used here.

We train the models by standard stochastic gradient descent with batch size $B=1024$ and a learning rate of $0.02$. 
We did not find consistent improvements for the hybrid models when applying an $L_2$ regularization and do not apply it in this work. We did find, however, a slight improvement for models containing only the pairwise energy $E_{pw}$, as explained in detail below.
We trained all models for $250$ epochs.

\section{Experimental Setting: Generating Distributions}
\label{sec:generator}
In this work, the experimental setting is given by a data generating distribution $p_G(s)\propto \exp(-E_G(s))$ over the configurations $\{-1,+1\}^N$, where $E_G(s)$ contains a pairwise part and an additional number of higher-order interactions:

\begin{align}
    E_G(s) &= E^G_{pw}(s) + E^G_{ho}(s) =  - \sum_{i<j} J^G_{ij} s_i s_j - \sqrt{\gamma} \sum_{I \in \mathcal{I}_G} \xi_I^{G} \prod\limits_{i \in I} s_i.
    \label{eq:generator}
\end{align}

$\mathcal{I}_G$ is a set of sets of indices determining the higher-order
interactions of the generator. Since we are interested in the effect of
additional higher-order interactions, we restrict ourselves to cases where $|I|
\geq 3$. In order to model the situation where a pairwise distribution is
probably a good approximation, we will keep these higher-order interactions
sparse and choose only a small subset of the  $~2^N$ possible interactions,
mostly only $N$.  The factor $\gamma$, which we call higher-order strength, is used to weight the two terms against each other (see below).
The specific interacting sets $I \in \mathcal{I}_G$ are independently and randomly chosen
either as only triplets or as interactions of order 3 to 10, according to the different settings we present in the following sections. 

The interaction parameters $\xi_I^G$ and the couplings $J_{ij}^G$ are independently sampled from
Gaussian distributions. In order to ensure that none of the two parts of the
generator completely dominates the distribution, we fine tune their relative strength for each sample as follows. For a system size of $N$, we generate Gaussian i.i.d couplings for the pairwise part of the generator, $J^G_{ij} \sim \mathcal{N}(0,1/N)$.
We call $\sigma^2_{G,pw}$ the variance of the induced pairwise energy across uniformly distributed configurations, $\sigma^2_{G,pw}= \mathrm{Var}[E^G_{pw}]=\sum_{i<j} (J^G_{ij})^2$.  Next, we generate i.i.d. parameters $\hat{\xi}^G_I \sim \mathcal{N}{(0,1)}$, compute the induced higher-order energy variance across uniformly sampled configurations, $\sigma^2_{G,ho}=\sum_{I} (\hat{\xi}^G_I)^2$, and finally
set $\xi^G_I = (\sigma_{G,pw}/\sigma_{G,ho})~\hat{\xi}^G_I$. 
We can then use $\gamma$ to set the ratio between the two variances: 
$\mathrm{Var}[E^G_{ho}]\,/\,\mathrm{Var}[E^G_{pw}]=\gamma$. We note that this procedure is not meant to balance the two terms perfectly for $\gamma=1$, but rather to give a well-defined starting point for the exploration of different values of $\gamma$. The idea of this work is to explore situations in which a pairwise model describes the variability in the generator well, but not perfectly. We therefore evaluate different values of $\gamma$ in terms of how it affects the training of a purely pairwise model on data from the generator and use this metric to decide which values of $\gamma$ are interesting.

We generate configurations independently sampled from the generator as follows. 
For $N<20$, it is feasible to calculate the probabilities involved
exactly. We therefore calculate the energies for all possible sequences,
exponentiate and normalize them, and then sample sequences using a standard
numeric library~\cite{harris2020array}. For larger $N$, we resort to the
standard Metroplis-Hastings algorithm, which we
parallelized on the GPU by running the energy evaluations on all sequences as one batch.  We used $N \cdot 10^4$ MC update steps for sampling.

\section{Results}
\label{sec:results}

\subsection{Reconstructing Pairwise Interactions with Neural Networks}
\label{sec:pairwise_training}

We analyse the effect that additional higher-order interactions in the
generating process might have on the reconstruction of the pairwise couplings by training the same models on data from generators with different higher-order strength $\gamma$. We call the criterion that we adopt to measure the reconstruction performance the reconstruction error $\epsilon$. It is a relative measure of the deviation of the inferred couplings $\hat{J}_{ij}$ from the true ones $J^G_{ij}$:

\begin{align}
    \epsilon = \sqrt{\frac{\sum\limits_{i<j} \left( J^{G}_{ij} - \hat{J}_{ij} \right)^2}{\sum\limits_{i<j} \left( J^{G}_{ij}\right)^2}}.
    \label{eq:reconstruction_error}
\end{align}

We expect that the additional interactions will have little
to no effect for small values of $\gamma$ in the generative model
\eqref{eq:generator}. In this case, we
can expect that training a purely pairwise model will lead to satisfactory results.
When increasing $\gamma$, however, the
generating distribution deviates significantly from a pairwise model,
and an increase in the reconstruction error can be expected using a purely pairwise model.  

For the experiments in this section, the generators contained $N$ uniformly sampled triplet interactions ($|I|=3\ \forall I \in \mathcal{I}_G$). Other details of the data generation process are given in Sec. \ref{sec:generator}, while the training procedure is the one outlined in Sec. \ref{sec:training}. 
We generated $M=5\cdot10^4$ training configurations for system size $N=64$, 
and $M=10^4$ for $N=16$.
The neural network part of the hybrid model, $E_{nn}$, was an MLP with one hidden layer of 128 units and $tanh$ activations. For the hybrid model, we evaluated both the explicit couplings in $E_{pw}$ and the reconstructed couplings obtained at the end of the training from Eq.~\eqref{eq:coupling_estimate}.

We compare the reconstruction based on the hybrid model against two other methods: The first is the commonly used regularized pseudolikelihood inference, which amounts to training only the pairwise part $E_{pw}$ of the hybrid model. In this setting, the possibilities of model mismatch and overfitting are often addressed by adding an $L_2$ regularization, which we therefore also add in our experiments for this model type. We found that a relatively low regularization strength $\lambda=0.01$ lead indeed to a slight improvement for a large range of $\gamma$ and used this value for all our experiments.

The second model we compare against is the energy-based model containing only the neural network part $E_{nn}$.

\begin{figure}
    \centering
    \includegraphics[width=1.00\textwidth]{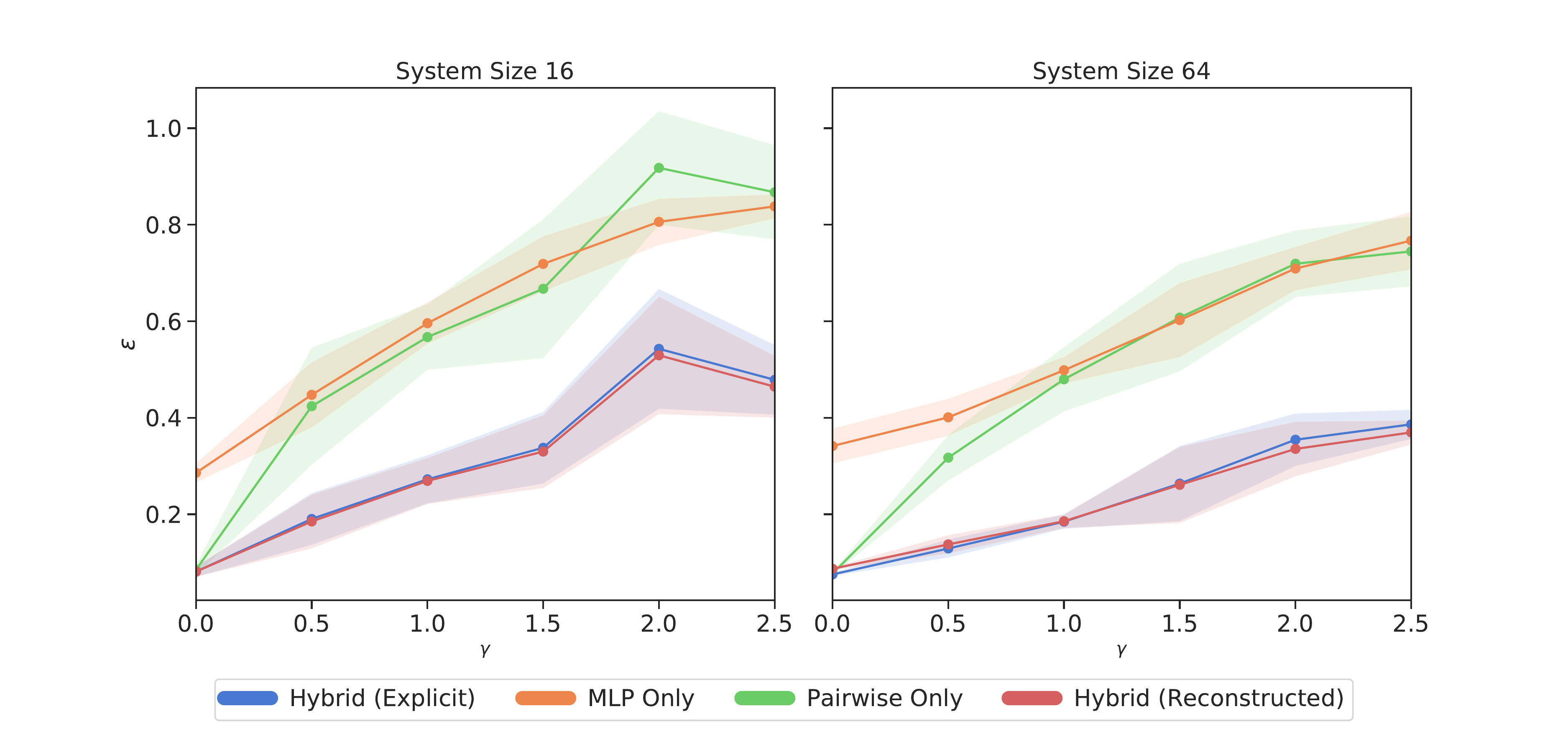}
    \caption{Reconstruction error for different system sizes $N$ and different models as a function of higher-order strength  $\gamma$ in the data generator. The data is generated by a pairwise model with $N$ additional interactions involving only 3 variables (see Eq. \eqref{eq:generator}). We show means and standard deviations over 5 runs. The reconstructed couplings for the Hybrid and the MLP only model are calculated using Eq.~\eqref{eq:coupling_estimate}. Both the hybrid and the MLP only model had a single layer of 128 hidden neurons.}
    \label{fig:effects_of_gamma_triplets}
\end{figure}

In Fig.~\ref{fig:effects_of_gamma_triplets} we show the error in the
inferred couplings with respect to the couplings in the generator.
While for all models the reconstruction degrades as $\gamma$ grows, the hybrid approach performs substantially better than models containing only the pairwise part $E_{pw}$ or only the neural network part $E_{nn}$. 
The explicit and the reconstructed couplings for the hybrid model yield similar result,
meaning that the learned $E_{nn}(s)$ function is approximately orthogonal to the pairwise family in this experiment. 
It is interesting to note that the neural network with 128 hidden neurons is insufficient to reconstruct the couplings. This confirms the idea that the
explicit pairwise model is useful in training. However, we will later show that
using networks with much larger capacity, the MLP only model can approach the hybrid model performance in some of the settings explored.

\subsection{Specificity of the Inferred Interactions}
Using the same experimental setting as in the previous section,
we investigate in detail how closely the trained hybrid model
matches the generator.

\begin{figure}
    \centering
    \includegraphics[width=1.0\textwidth]{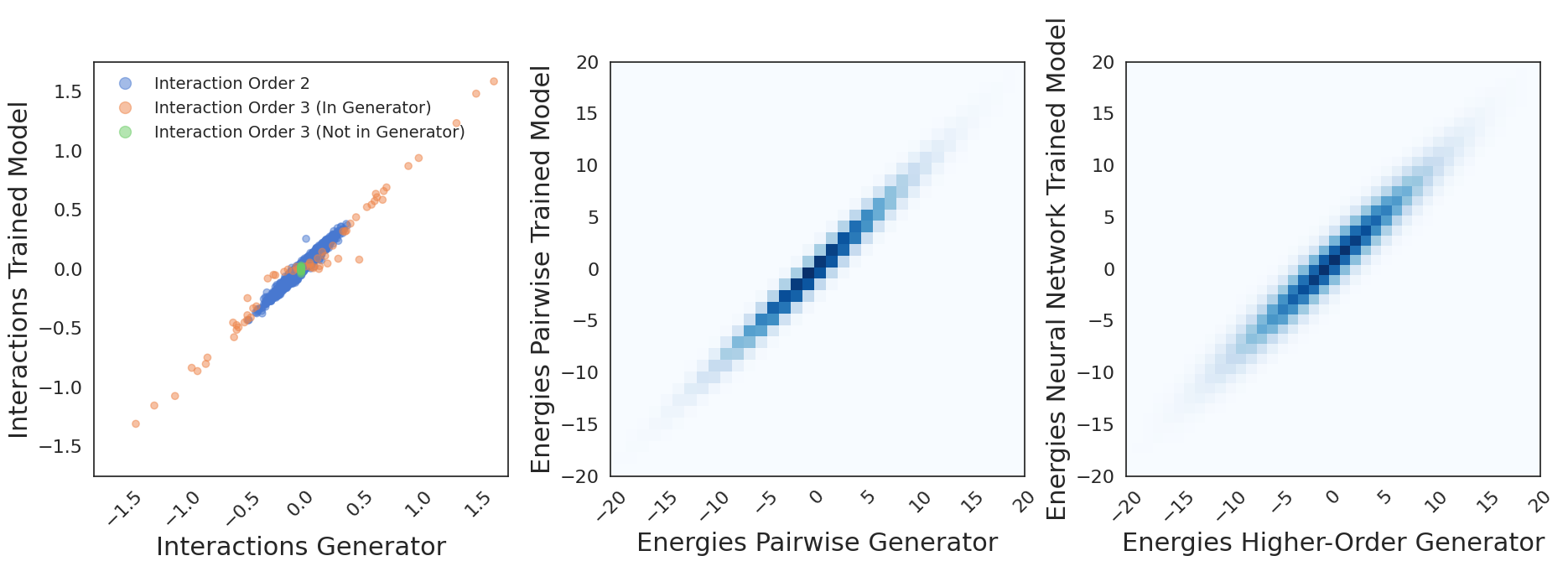}
    \caption{Inferred versus true interactions for system size $N=64$. The generator includes $N$ triplet interactions and $\gamma$ is set to $1.0$. (Left) Blue points refer to pairwise interactions, orange points to all $64$ triplet interactions present in the generator and green to $64$ random triplet interactions not present in the generator. (Center and Right) Relation of the energies between the submodels of the generator (pairwise and higher-order) and the trained model (pairwise and neural network). The color intensity is proportional to the density of points. The hybrid model contained a single hidden layer with $128$ hidden neurons. All interactions were estimated using Eq.~\eqref{eq:interaction_estimate} using $10^6$ samples.}
    \label{fig:histogram_triplets64}
\end{figure}

In Fig.~\ref{fig:histogram_triplets64} (left) we compare the reconstructed interaction parameters from the hybrid model through Eqs. \eqref{eq:interaction_estimate} and \eqref{eq:coupling_estimate} to the corresponding ones in the generator.
The interaction parameters that we estimate are all pairwise interactions, the $N$
triplet interactions that are present in the generator and $N$ random triplet
interactions not present in the generator. 
Pairwise interactions are well fitted, as well as the strongest triplet interactions in the generator. Some weaker triplet interactions in the generator are underestimated instead. The triplet interactions not contained in the generator are close to $0$ in the hybrid model. These results indicate that the hybrid model does not only learn an effective model of the generator, but extracts the true interactions in the underlying system.

In Fig.~\ref{fig:histogram_triplets64} (center and right) we show that the energies calculated from the
pairwise part in the generator are strongly correlated with the energies from
the pairwise part in the trained model and the energies calculated from the
trained neural network are strongly correlated with the energies coming from
the higher-order interactions in the generator. 

\begin{figure}
    \centering
    \includegraphics[width=1.0\textwidth]{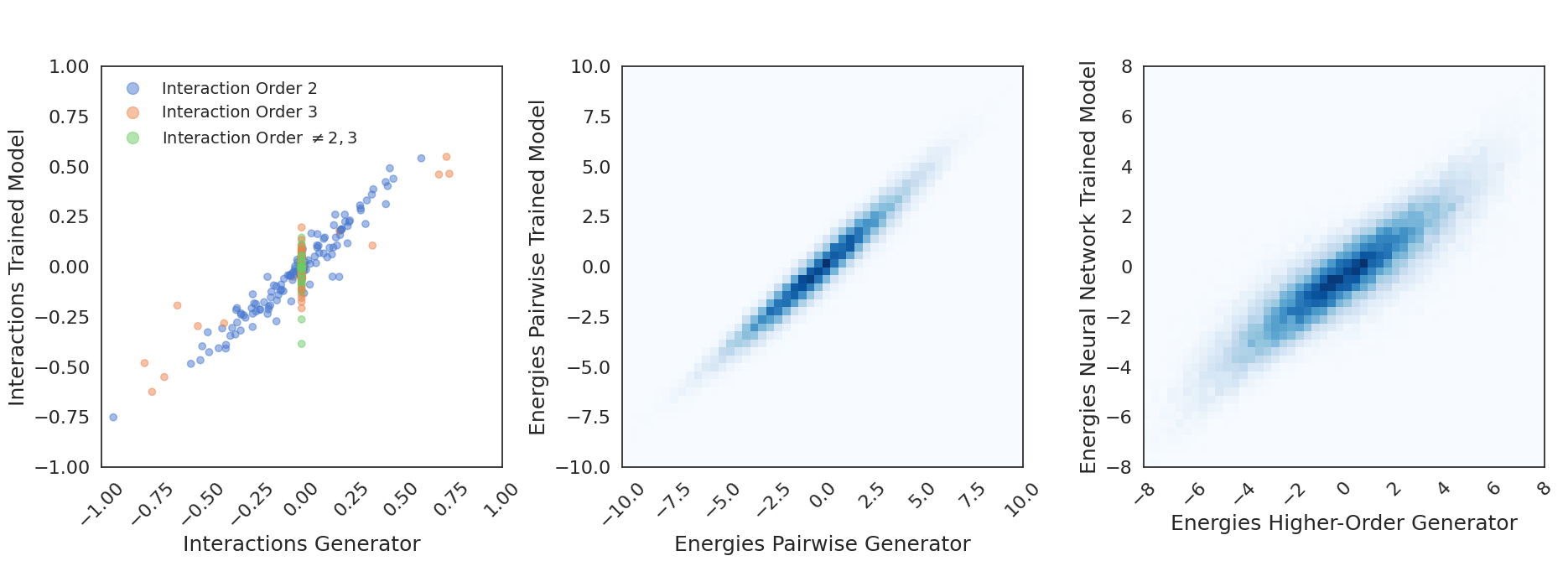}
    \caption{Inferred versus true interactions for system size $N=16$. The generator included $N$ triplet interactions and $\gamma$ was set to $1.0$ and the hybrid model had a single hidden layer with $128$ hidden units. The left panel shows all interactions color coded by their order: blue points refer to pairwise interactions, orange points to triplet interactions and green to all other orders. The middle and right panel show the relation of the energies between the submodels of the generator (pairwise and higher-order) and the trained model (pairwise and neural network). The color is proportional to pairs of energy values that fall in the corresponding quadrant. All interactions were recovered using Eq.~\eqref{eq:interaction_estimate} using all possible sequences and should be exact. The energies in the middle and right panel also correspond to all possible sequences.}
    \label{fig:histogram_triplets}
\end{figure}

See also Supplementary Fig.~\ref{fig:histogram_triplets} for the same experiment on a smaller system.

\subsection{Varying Neural Network Sizes}
\begin{figure}
    \centering
    \includegraphics[width=0.8\textwidth]{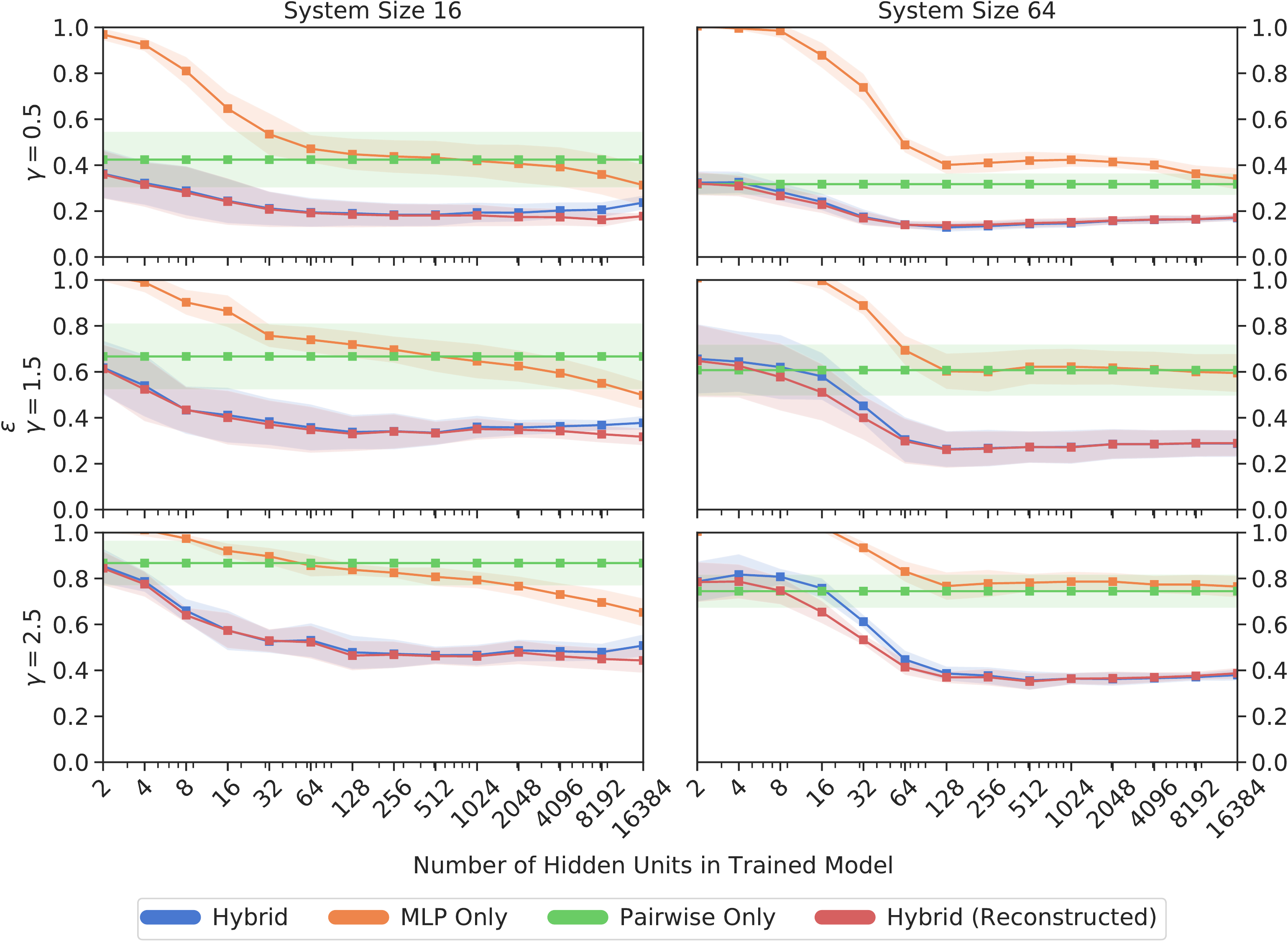}
    \caption{Reconstruction error of couplings in presence of triplet interactions in the generator, for varying number of hidden neurons in the trained model and different values of $\gamma$. We used $M=10^4$ training samples for $N=16$ and $M=5 \cdot 10^4$ training samples for $N=64$. Shown are means and standard deviation over 5 independent samples.}
    \label{fig:vary_h_triplets}
\end{figure}

In order to evaluate the impact of the neural architecture used in the hybrid model \eqref{eq:hybrid}, 
we repeat the
experiments with different sizes for the hidden layer of the MLP. As in the previous section, we
keep the higher-order interactions in the generator restricted to $N$ triplets, where $N$ is the system size. We vary the number of hidden
neurons between $2$ and $16384$ in powers of $2$. The results in
Fig.~\ref{fig:vary_h_triplets} indicate that size of the neural network
has only a small effect on the error above a certain threshold (around $128$ in
this specific case). While using a pure pairwise model for training leads
to a quickly increasing reconstruction error (as already visible in
Fig.~\ref{fig:effects_of_gamma_triplets}), the addition of a single layer neural
network with even a small number of hidden neurons (on the order of the system
size $N$) leads to a significantly better reconstruction of the pairwise couplings in the generator.

Varying the number of hidden neurons allows us also to test the hypothesis that a
sufficiently large neural network on its own is enough for inferring the
pairwise couplings. In this setting, the models containing only an MLP  approach the performance of the hybrid model only for $N=16$ and for very wide networks, while a large gap remains at $N=64$. We note that where models based only on a neural network perform well in terms of the reconstruction error, 
the hybrid model obtains comparable results with two orders of magnitude
less parameters. 
It is also to be said, however, that this comparison is not completely fair since
the hybrid model contains an inductive prior by design,
which the pure neural network model lacks. Still, we take this observation as evidence that adding a pairwise part in the trained model is sensible if the generating distribution is expected to contain a significant pairwise part.

\subsection{Varying the Interaction Orders in the Generator}
\label{sec:higher_order_mess}

In the preceding sections we restricted ourselves to triplet interactions in
the generator. In order to probe the limits of our approach, we repeat the
experiments with generators that contain $N$ higher-order interactions up to order $10$, leaving
all other characteristics like training set size and training approach the
same. The order of each  interaction is chose from a uniform distribution between $3$ and $10$, and the variables involved in each interaction are a random subset of all variables. 

We note that this is a very ambitious test: our hope is that the neural
network picks up the higher-order interactions in the generator, which are of
the type $\xi \prod_{i=1}^{I} s_i$, where $I$ is the interaction order and
$\xi$ the corresponding parameter. This means that we
try to fit a combination of overlapping sparse parity problems of up to $10$
inputs. While constructing a solution to a single instance of such a
problem is easy using a single hidden layer with continuous weights (see
e.g.~\cite{franco2001generalization}), parity functions are generally
considered among the hardest functions to
learn from data~\cite{tesauro1988scaling}. While we might be able to
alleviate this problem by adding more layers to the neural network, we consider
this to be out of scope for the current work and note that in a realistic
application the size of the underlying interactions is often not known. Even in
this hard case, however, one could expect that the neural network gives a
contribution to the quality of training by fitting at least some of the
variability due to the higher-order interactions.

\begin{figure}
    \centering
    \includegraphics[width=1.0\textwidth]{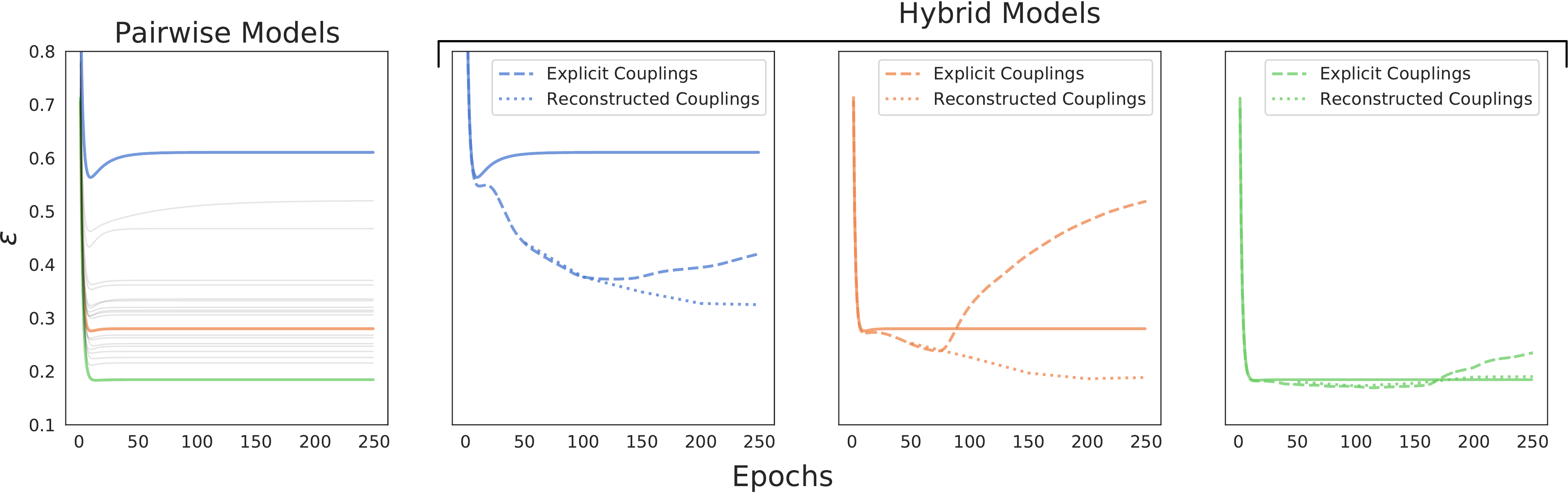}
    \caption{Reconstruction error of couplings in presence of higher-order (3 to 10) interactions in the generator as a function of the training epoch. The higher-order interaction strength $\gamma$ was set to $1.5$, the system size is $N=64$, and we used $M=5\cdot 10^4$ training samples. (Left) Reconstruction error for $20$ independent systems using only a pairwise model for training. The system corresponding to the colored lines are also used in the 3 right panels. (Right 3 panels) Reconstruction error given by pairwise only models (solid lines), and by hybrid models using either explicit or reconstructed couplings. The hybrid models contained a single hidden layer with $256$ hidden units. 
    \label{fig:higher_order_specifics}} 
\end{figure}

While also in this setting we report generally better performance of  the hybrid approach over the pairwise only and neural network only approaches, the gain is not as large as in the case of triplet interactions of the last sections (see Supplementary Fig. \ref{fig:higher_order_mess}). 
Moreover, in this setting the explicit couplings of the hybrid model significantly deviate
from the couplings reconstructed using Eq. \ref{eq:coupling_estimate} at the end of training, as can be seen
in Fig. \ref{fig:higher_order_specifics}. While the additional reconstruction step is computationally cheap, these observations suggest that additional constraints for keeping the pairwise interactions in the neural network small might lead to further improvements. 
In Supplementary Section \ref{sec:absorber} we present a rough way of doing this and speculate about more sophisticated approaches.

\section{Discussion}

In this work we have shown that adding neural networks to pairwise models can
improve the quality of reconstruction of pairwise interactions if the distribution
underlying the data generating process contains additional higher-order
interactions, as it typically occurs in natural data. 
While both the explicitly pairwise part and the neural network 
part of the hybrid model may contribute to the reconstructed couplings in general, we showed that in certain settings the neural network and the pairwise model specialize in fitting the separate parts of the generating model.

There are many directions for future investigations. Systematic exploration of
the neural architecture employed, which we did not pursue at great length in
this work, could yield significant improvements.  Different training
methods for energy based models could be applied, possibly speeding up simulations or giving more robust predictions. We also did not check the quality of the trained models when used as generative distributions, which might be an important factor when applying similar methods for example to protein design. In addition, constraining the neural network to account only for higher-order interactions in a more sophisticated way might lead to further improvements. 

To the best of our knowledge, this work is the first one that solves the inverse problem by using the couplings reconstructed from a neural network. This leads to another line of possible research, were the training of possibly very large and complex generative models without explicit pairwise couplings is followed by a reconstruction step.  In principle, common architectures like GANs or autoregressive networks could be adapted at little additional computational cost.

The next immediate step, however, would be to screen the current application
domains of pairwise models and translate the improvements observed in the
well-controlled settings in this work to real-world data. Many of these fields
present idiosyncrasies, for example in the data characteristics, the expected
topology of the underlying interactions or the additional tricks in training
or preprocessing that are important for achieving good results when using
purely pairwise models. We therefore expect that some additional adaptions to
our method are necessary in these cases. Given, however, that in most or all
applications pairwise models are used as effective models and one would expect
higher-order interactions to play a role in almost all complicated real-world
scenarios, we believe that our work presents a very promising perspective.

\bibliographystyle{unsrt}
\bibliography{references}

\newpage
\FloatBarrier

\beginsupplement{}
\section*{Supplemental Material}

\subsection{Using Pseudolikelihoods for training EBMs}
\label{sec:pl_sup}

Pseudolikelihoods are often used as an alternative to an intractable or
at least computationally expensive likelihood
~\cite{besag1977efficiency}. It has been applied successfully to pairwise
models~\cite{hyvarinen2006consistency, aurell2012inverse, ekeberg2013improved, Decelle2014}. We show here how it can
be applied to a generic Energy-Based Model, and add some  considerations specific to pairwise models. We note that while maximum pseudolikelihood is a  widely applied method for training simple Energy-Based Models, to the best of our knowledge this is the first time it has been used for  training deep feed-forward neural networks. 

We assume the data that we want to model to consist of configurations $(s_1, \ldots,
s_N)$ of categorical variables of length $N$ and we will use $q$ to denote the
number of categories. A common method for fitting a probability distribution
$p_{\Theta}(s)$ with parameters $\Theta$ to a training set of sequences
$\{s^m\}_{m=1}^M$ is to find the $\Theta^{*}$ for which 

\begin{align}
    \Theta^{*} = \underset{\Theta}{\mathrm{argmax}}~\sum\limits_{m=1}^M \log p_{\Theta}(s^m),
\end{align}

which corresponds to a \textit{maximum-likelihood} solution. For an Energy-Based Model (EBM) $p_{\Theta}(s) = \frac{e^{-E_\Theta(s)}}{Z_{\Theta}}$, where $E_{\Theta}(s)$ is the energy function, this would correspond to solving

\begin{align}
    \Theta^{*} = \underset{\Theta}{\mathrm{argmax}}~\frac{1}{M}\sum\limits_{m=1}^M \left[ -E_{\Theta}(s^m) - \log Z_{\Theta} \right],
\end{align}

for example by gradient descent methods. The general problem in this approach is that the normalization constant $Z_{\Theta} = \sum\limits_{s} e^{-E_\Theta (s)}$, where we sum over all possible configurations $s$, contains $q^N$ terms. This is intractable even for modest $N$ and in the case of binary variables, where $q=2$. The idea of pseudolikelihoods is to replace the likelihood objective by

\begin{align}
    \Theta^{*} = \underset{\Theta}{\mathrm{argmax}}~\frac{1}{M}\sum_{i=1}^{N}\sum\limits_{m=1}^M \log p_{\Theta}\left(s^m_i | s^m_{/i}\right),
\end{align}

where $p_{\Theta}\left(s^m_i | s^m_{/i}\right)$ is the probability of symbol $s^m_i$ in sequence $m$, given the other symbols. We therefore train the distribution by using it for predicting a \textit{missing} symbol from the other symbols.

Other variations are possible, for example to discard the sum over $i$ and find
a maximum set of $\Theta_i^*$ for every $i$ independently. We found the
approach with the sum to be conceptually easier and in the applications known
to us, the performance seems to be the same~\cite{ekeberg2013improved}. While
it can be shown that this new objective has the same maximum as the original
likelihood under certain conditions \cite{mozeika2014consistent}, this is for example not generally true if the
training samples come from a different model class than $p_{\Theta}$, which is
true in our case. In this work, we are interested in whether we can make
training using this objective work in practice and refrain from further
theoretical analysis.We note that we have not restricted the form of $E_{\Theta}$. In the models we
analyse in this work, the energy is calculated by a sum of the energy of a pairwise model and a neural network.

Neglecting the sum over $i$ and $m$ for the time being, we can write the quantity $\log p_{\Theta}(s_i | s_{/i})$ for an EBM as 

\begin{align}
    \log p_{\Theta}(s_i | s_{/i}) = \log \frac{p_{\Theta}(s)}{p_{\Theta}(s_{/i})} = \log \frac{p_{\Theta}(s)}{\sum_{\hat{s}_i=1}^q p_{\Theta}(\hat{s}_i, s_{/i})},
\end{align}

where we used the notation $(\hat{s}_i, s_{/i})$ for the configuration $s$ after $s_i$ has been replaced with $\hat{s}_i$. Since the normalization constant $Z_{\Theta}$ appears in the both the numerator and denominator, it cancels and we are left with

\begin{align}
    \log p_{\Theta}(s_i | s_{/i}) = \log \frac{e^{-E_{\Theta}(s)}}{\sum_{\hat{s}_i=1}^{q} e^{-E_{\Theta}(\hat{s}_i, s_{/i})}} = -\log \left( 1 + \sum_{\hat{s}_i \neq s_i} e^{E_{\Theta}(s)-E_{\Theta}(\hat{s}_i, s_{/i})}\right)
    \label{eq:pl_log}
\end{align}

The sum in this expression can be computed efficiently, using $q$ evaluations
of $E$. This means that including the sum over $i$ and replacing the sum over
$m$ with a sum over a mini-batch of $B$  in a \textit{stochastic gradient
descent} (SGD) setting, we need $q \cdot N \cdot B$ evaluations of $E$ for a
single gradient step, corresponding to $q \cdot N$ forward passes.

In the case of binary strings with $s_i \in \{\pm 1\}$ and a pairwise model
$E_{\Theta}(s) = -\sum_{i<j} J_{ij} s_k s_j$ with parameters $\Theta \equiv J$, we can
simplify further by noticing that

\begin{align}
    E(s) - E(\hat{s}_i, s_{/i}) = (\hat{s}_i - s_i) \sum\limits_{j\neq i} J_{ij} s_j  ,
\end{align}
where we identified $J_{ij}=J_{ji}$ for convenience. 
Since in Eq.~\eqref{eq:pl_log} we sum only over $\hat{s_i} \neq s_i$ and in this case $\hat{s}_i - s_i = -2 s_i$, this leads to

\begin{align}
    \log p_{\Theta}(s_i | s_{/i}) = -\log \left( 1 + e^{-2 s_i F_i \left( J, s_{/i} \right)} \right),
\end{align}

where $F_i(J, s_{/i}) = \sum\limits_{j\neq i} J_{ij} s_j$. This means that in this model
class, we do not need to evaluate the full energy, which contains $\Theta(N^2)$
terms, but only the part of the energy involving the variable $s_i$, which
contains only $\Theta(N)$ terms. The quantities $F_i \left( J, s_{/i}
\right)$ can  be obtained for a whole batch of sequences using
matrix multiplication, which is very efficient on modern GPUs.

\subsection{Absorbing Pairwise Interactions from the Neural Network}
\label{sec:absorber} 

During training, we did not enforce a division of labour between the two parts of the
hybrid models, which means that the neural network is not discouraged in any
way from fitting also pairwise interactions.  While extracting the pairwise
coefficients from the entire hybrid model and constructing an effective
pairwise model is a way of solving this after training, it would be more
satisfactory to include this also in the training procedure.  The
cleanest way of ensuring only higher-order interactions in the neural network
would be to constrain the optimization of the neural network to the part of
parameters space where it does not contain pairwise interactions. 
In practice, Eq. \eqref{eq:interaction_estimate} could be used to create a regularization term penalizing all
pairwise interactions:

\begin{align}
    \frac{1}{N^2}\sum_{ij}\left(\mathbb{E}_s\ s_i s_j\, E_{nn}(s)\right)^2 = \mathbb{E}_{s,s'}\ E_{nn}(s) E_{nn}(s')\,q^2(s,s'), 
\end{align}
where the expectation is over uniformly sampled Ising configurations and 
$q(s,s')=\frac{1}{N}\sum_i s_i s'_i$ is the overlap between two configurations.
This expression can be approximately evaluated by Monte Carlo sampling. 
While this approach seems promising, we did not pursue it in this exploratory analysis. 

A different approach instead is to counter
the pairwise interactions in the neural network by using an additional pairwise model. To this end, we define a
new energy

\begin{align}
    E(s) = E_{pw}(s) + E_{nn}(s) - \hat{E}_{pw}(s).
    \label{eq:helper}
\end{align}

Here, $E_{pw}$ and $E_{nn}$ are the same as in the hybrid models of the
preceding sections. The new term $\hat{E}_{pw}$ is another pairwise model, but
it is excluded from the gradient descent step and we set its couplings
explicitly every $k$ epochs. The values of these couplings are the pairwise
interactions extracted from $E_{nn}$ using Eq~\eqref{eq:interaction_estimate}.
The idea is to estimate the pairwise terms in the expansion of the neural
network energy $E_{nn}$ and \textit{absorb} these interactions in the
additional $\hat{E}_{pw}$, which we therefore call an absorber model. After
setting the couplings of this absorber, the last two terms on the right hand
side of Eq.~\eqref{eq:helper} should contain approximately no pairwise
interactions, i.e.

\begin{align}
    \sum_{s} s_i s_j  \left(E_{nn}(s) - \hat{E}_{pw}(s)\right) \approx 0
\end{align}

for all $i,j$. This leaves the term $E_{pw}$ as the only one with significant
pairwise interactions. While we could do this in principle after every epoch or
even after every gradient step, this would make the computations unfeasibly
slow since at every step we estimate the pairwise interactions in $E_{nn}$
using $10^6$ samples. We
therefore restrict ourselves to doing the estimate less frequently, every $k=5$
epochs in our experiments.

\begin{figure}
    \centering
    \includegraphics[width=1.0\textwidth]{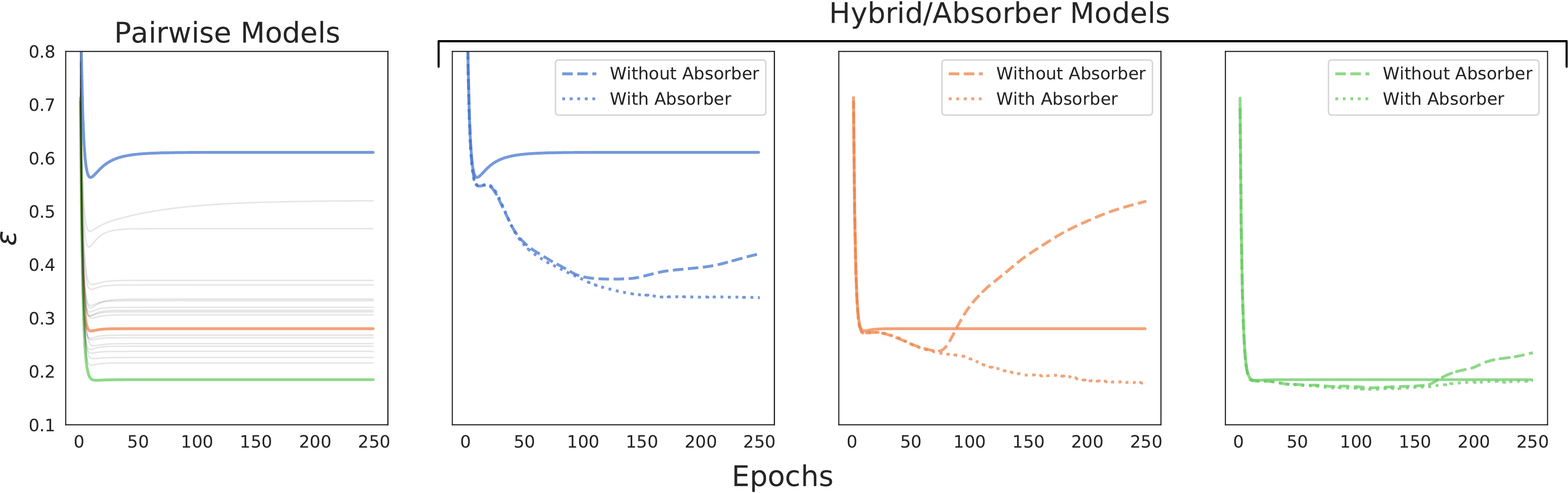}
    \caption{Reconstruction error with $64$ higher-order (3 to 10) in generator for $\gamma=1.5$ and trained with $M=5\cdot 10^4$ samples. The training samples in this figure are the same as in Fig.~\ref{fig:higher_order_specifics}. Left panel: Reconstruction error for $20$ independent runs using only a pairwise model for training. The training samples corresponding to the colored lines were further used as a training set for the hybrid and absorber models shown in the 3 right panels. Right 3 panels: Reconstruction error for trained models containing only pairwise terms (solid lines), reconstruction error for hybrid models (dashed lines) and reconstruction error for hybrid models with absorber terms (dotted lines)}

    \label{fig:absorber} 
\end{figure}

In Fig.~\ref{fig:absorber} it can be seen that using these additional absorbers
improves the training of the couplings significantly. We used the same
training samples as for Fig~\ref{fig:higher_order_specifics} and also left the
other training characteristics the same. The results are very similar to what
would have been obtained by reconstructing the couplings at every step (compare
also Fig.~\ref{fig:higher_order_specifics}). While this also means that there
was no strong improvement over approach of reconstructing the couplings after the training has ended, we think it still
noteworthy that enforcing that the pairwise model $E_{pw}$ should be the one
solely responsible for fitting pairwise interactions is possible during
training.

\FloatBarrier{}
\subsection{Additional Figures}
\label{sec:additional}

\begin{figure}[H]
    \centering
    \includegraphics[width=1.0\textwidth]{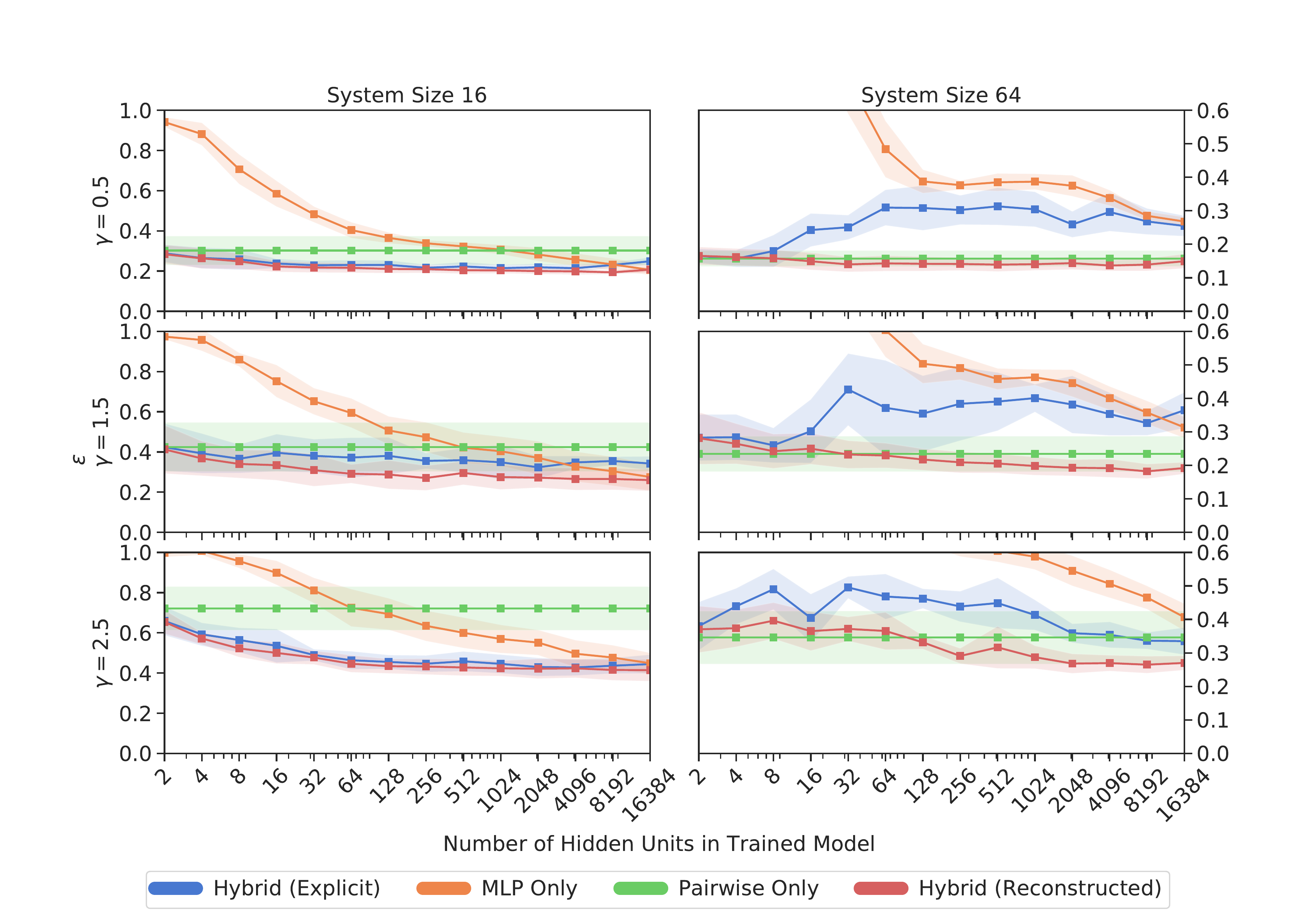}
    \caption{Pairwise reconstruction errors with higher-order (size 3 to 10) interactions in generator for varying number of hidden neurons in the trained model and 3 different values of $\gamma$. The number of these interactions is equal to $N$. Training was done with $M=10^4$ samples for $N=16$ and $M=5 \cdot 10^4$ samples for $N=64$. Shown are means and standard deviation over 5 runs. The reconstruction error is defined in Eq.~\eqref{eq:reconstruction_error}. The blue line corresponds to reconstruction error calculated using the couplings of the pairwise part of the hybrid model, the red line to the pairwise interactions reconstructed from the complete hybrid model using Eq. \eqref{eq:coupling_estimate}}~\label{fig:higher_order_mess}.
\end{figure}

\end{document}